\documentstyle[aps,preprint,tighten,floats,psfig]{revtex}

\begin{document}
\draft
\preprint{\vbox{\it Submitted to Phys. Lett. B \hfill\rm CU-NPL-1147}}

\title{Predicative Ability of QCD Sum Rules for Decuplet Baryons}
\author{Frank X. Lee}
\address{Nuclear Physics Laboratory, Department of Physics,
University of Colorado, \\Boulder, CO 80309-0446}
\date{\today}
\maketitle

\begin{abstract}
QCD sum rules for decuplet baryon two-point functions are 
investigated using a comprehensive Monte-Carlo based procedure.
In this procedure, all uncertainties
in the QCD input parameters are incorporated simultaneously, 
resulting in realistic estimates of the uncertainties 
in the extracted phenomenological parameters.
Correlations between the QCD input parameters and the phenomenological
parameters  are studied by way of scatter plots. 
The predicted couplings are useful in evaluating matrix elements 
of decuplet baryons in the QCD sum rule approach.
They are also used to check a cubic scaling law between 
baryon couplings and masses, as recently found by Dey and 
coworkers. The results show a significant reduction in the 
scaling constant and some possible deviations from the cubic law.
\end{abstract}
\vspace{1cm}

\parskip=2mm
\section{Introduction}
\label{intro}

	The QCD sum rule method~\cite{SVZ79} is a powerful tool in
revealing the deep connection between hadron phenomenology and 
QCD vacuum structure via a few condensate parameters 
--- vacuum expectation values of QCD local operators.
This nonperturbative method has been successfully applied to a variety of 
problems to gain a field-theoretical understanding 
into the structure of hadrons (for a review
of the early work, see Ref.~\cite{RRY85}), and continues to be 
an active field~\cite{archive}. 

The decuplet baryons have been studied in the past using the QCD 
sum rule method. In Ref.~\cite{Bely82}, $\Delta$ was first studied.
It was later extended to $\Sigma^*$ in Ref.~\cite{Bely83}.
In Ref.~\cite{RRY82}, sum rules for the decuplet family were given  
(they were also reported in Ref.~\cite{RRY85}). 
Ratio method was used there to extract the masses. 
No attempt was made to extract the current couplings. 
No anomalous dimension corrections were considered.
In Ref.~\cite{Chung84}, $\Delta$ was studied including leading-order 
$\alpha_s$ corrections. In Ref.~\cite{Derek90}, $\Delta$ was studied.
A perusal of these works reveals discrepancies among the QCD sum rules 
derived where comparisons are possible. 
The analysis methods employed were relatively crude.
Often a 10\% or better accuracy was claimed in all the extracted 
quantities without the support of rigorous error analysis. 

In this work, we decide to re-derive the sum rules for all members 
of the decuplet family, consistently including operators
up to dimension 8, first order strange quark mass corrections, 
flavor symmetry breaking of the strange quark condensates,
anomalous dimension corrections, 
and factorization violation of the four-quark condensate. 
Furthermore, we try to assess quantitatively
the errors in the phenomenological parameters,
using a Monte-Carlo based procedure~\cite{Derek96}. 
This procedure incorporates all uncertainties in the 
QCD input parameters simultaneously, and translates them into 
uncertainties in the phenomenological parameters, with careful regard to 
OPE convergence and ground state dominance.
The goal is to get a realistic understanding of the predicative ability
of the standard implementation of the QCD sum rule approach for 
the decuplet baryon two-point functions.
Of particular interest is the baryon coupling to its current.
This quantity is crucial to studying matrix elements of these 
baryons, such as magnetic moments, transition moments, axial charges, 
tensor charges, etc., because they utilize the coupling as normalization.

\section{Method}
\label{elem}

The starting point is the two-point correlation function 
in the QCD vacuum
\begin{equation}
\Pi_{\scriptscriptstyle \mu\nu}(p)=i\int d^4x\; e^{ip\cdot x}\langle 0\,|\,
T\{\;\eta_{\scriptscriptstyle \mu}(x)\,
\bar{\eta}_{\scriptscriptstyle \nu}(0)\;\}\,|\,0\rangle,
\label{cf2pt}
\end{equation}
where $\eta_{\scriptscriptstyle \mu}$ is the interpolating field (or current)
with the quantum numbers of the baryon in question.
The interpolating field excites the ground state 
as well as the excited states of the baryon 
from the QCD vacuum. The ability of the interpolating field to 
annihilate the {\em ground state} baryon into the QCD vacuum 
is described by a phenomenological parameter $\lambda_B$ (called current
coupling or pole residue)
\begin{equation}
\langle 0\,|\,\eta_{\scriptscriptstyle \mu}\,|\,Bps\rangle
=\lambda_B\,u_{\scriptscriptstyle \mu}(p,s),
\end{equation}
where $u_{\scriptscriptstyle \mu}$ is the Rarita-Schwinger spin-vector.
This parameter plays an important role in evaluating 
matrix elements of the baryons.

The lowest dimensional interpolating fields for the decuplet 
are uniquely defined.
Assuming SU(2) symmetry in the u and d quarks, they can be written as:
\begin{equation}
\eta_{\scriptscriptstyle \mu}^{\scriptscriptstyle \Delta}(x)=
\epsilon^{abc}\left(u^{aT}(x)C\gamma_\mu u^b(x)\right) u^c(x),
\end{equation}
\begin{equation}
\eta_{\scriptscriptstyle \mu}^{\scriptscriptstyle {\Sigma^*}}(x)=
\sqrt{1/3}\epsilon^{abc}
\left[2\left(u^{aT}(x)C\gamma_\mu s^b(x)\right) u^c(x)
+\left(u^{aT}(x)C\gamma_\mu u^b(x)\right) s^c(x)\right],
\end{equation}
\begin{equation}
\eta_{\scriptscriptstyle \mu}^{\scriptscriptstyle {\Xi^*}}(x)=
\sqrt{1/3}\epsilon^{abc}
\left[2\left(s^{aT}(x)C\gamma_\mu u^b(x)\right) s^c(x)
+\left(s^{aT}(x)C\gamma_\mu s^b(x)\right) u^c(x)\right],
\end{equation}
\begin{equation}
\eta_{\scriptscriptstyle \mu}^{\scriptscriptstyle \Omega}(x)=
\epsilon^{abc}\left(s^{aT}(x)C\gamma_\mu s^b(x)\right) s^c(x),
\end{equation}
Here $C$ is the charge conjugation operator and the superscript $T$ 
means transpose.

The QCD sum rules are derived by calculating the correlator
in~(\ref{cf2pt}) using Operator-Product-Expansion (OPE), on the one
hand, and matching it to a phenomenological representation, on the
other. The obtained tensor structure has the form
\begin{equation}
\Pi_{\scriptscriptstyle \mu\nu}(p)=
\Pi_1(p^2)\;g_{\scriptscriptstyle \mu\nu}
+\Pi_2(p^2)\;g_{\scriptscriptstyle \mu\nu}\hat{p} +\cdots,
\end{equation}
where the hat notation denotes $\hat{p}=p^\alpha\,\gamma_\alpha$.
Two QCD sum rules can be derived from the two invariant functions 
$\Pi_1(p^2)$ and $\Pi_1(p^2)$ for each member using standard procedure.
The sum rule from $\Pi_1$ is called chiral-odd
since it involves dimension-odd condensates only. 
Similarly, the sum rule from $\Pi_2$ is called chiral-even since it 
involves dimension-even condensates only. 
These two structures are considered because they receive contributions 
from spin-3/2 states only.
The chiral-odd sum rules at the structure 
$g_{\scriptscriptstyle \mu\nu}$ are given 
for $\Delta$:
\begin{equation}
{4\over 3}\,a\; E_1\,L^{16/27}\;M^4
- {2\over 3}\,m^2_0 a\; E_0\,L^{2/27}\;M^2
- {1\over 18}\,a\,b\; L^{16/27}
=\tilde{\lambda}_{\scriptscriptstyle \Delta}^2\;
M_{\scriptscriptstyle \Delta}\;e^{-M_{\scriptscriptstyle \Delta}^2/M^2},
\label{Delta1}
\end{equation}
for ${\Sigma^*}$:
\begin{eqnarray}
& &
{4\over 9}\,(f_s+2)\,a\; E_1\,L^{16/27}\;M^4
-{2\over 9}\,(f_s+2)\,m^2_0 a\; E_0\,L^{2/27}\;M^2
- {1\over 54}\,(f_s+2)\,a\,b\; L^{16/27}
\nonumber \\ & &
+{1\over 2}\,m_s\; E_2\,L^{-8/27}\;M^6
+ {2\over 3}\,m_s\,\kappa_v a^2\;L^{16/27}
=\tilde{\lambda}_{\scriptscriptstyle {\Sigma^*}}^2\;
M_{\scriptscriptstyle {\Sigma^*}}\;e^{-M_{\scriptscriptstyle {\Sigma^*}}^2/M^2},
\label{Sigma1}
\end{eqnarray}
for ${\Xi^*}$:
\begin{eqnarray}
& &
{4\over 9}\,(2f_s+1)\,a\; E_1\,L^{16/27}\;M^4
-{2\over 9}\,(2f_s+1)\,m^2_0 a\; E_0\,L^{2/27}\;M^2
- {1\over 54}\,(2f_s+1)\,a\,b\; L^{16/27}
\nonumber \\ & &
+m_s\; E_2\,L^{-8/27}\;M^6
+ {4\over 3}\,m_s\,f_s\,\kappa_v a^2\;L^{16/27}
=\tilde{\lambda}_{\scriptscriptstyle {\Xi^*}}^2\;
M_{\scriptscriptstyle {\Xi^*}}\;e^{-M_{\scriptscriptstyle {\Xi^*}}^2/M^2},
\label{Xi1}
\end{eqnarray}
and for $\Omega$:
\begin{eqnarray}
& &
{4\over 3}\,f_s\,a\; E_1\,L^{16/27}\;M^4
- {2\over 3}\,f_s\,m^2_0 a\; E_0\,L^{2/27}\;M^2
- {1\over 18}\,f_s\,a\,b\; L^{16/27}
\nonumber \\ & &
+{3\over 2}\,m_s\; E_2\,L^{-8/27}\;M^6
+ 2\,m_s\,f^2_s\,\kappa_v \,a^2\;L^{16/27}
=\tilde{\lambda}_{\scriptscriptstyle \Omega}^2\;
M_{\scriptscriptstyle \Omega}\;e^{-M_{\scriptscriptstyle \Omega}^2/M^2}.
\label{Omega1}
\end{eqnarray}
The chiral-even sum rules
at the structure $g_{\scriptscriptstyle \mu\nu}\hat{p}$ are given 
for $\Delta$:
\begin{equation}
{1\over 5}\; E_2\,L^{4/27}\;M^6
- {5\over 72}\,b\; E_0\,L^{4/27}\;M^2
+ {4\over 3}\,\kappa_v a^2\;L^{28/27}
- {7\over 9}\,m^2_0 a^2\; L^{14/27}{1\over M^2}
=\tilde{\lambda}_{\scriptscriptstyle \Delta}^2\;
e^{-M_{\scriptscriptstyle \Delta}^2/M^2},
\label{Delta2}
\end{equation}
for ${\Sigma^*}$:
\begin{eqnarray}
& &
{1\over 5}\; E_2\,L^{4/27}\;M^6
- {5\over 72}\,b\; E_0\,L^{4/27}\;M^2
+ {4\over 9}\,(2f_s+1)\,\kappa_v a^2\;L^{28/27}
- {7\over 27}\,(2f_s+1)\,m^2_0 a^2\; L^{14/27}{1\over M^2}
\nonumber \\ & &
+{1\over 3}\,m_s\,(4-f_s)\, a\; E_0\,L^{4/27}\;M^2
-{1\over 18}\,m_s\,(14-5f_s)\,m^2_0 a\;L^{-10/27}
=\tilde{\lambda}_{\scriptscriptstyle {\Sigma^*}}^2\;
e^{-M_{\scriptscriptstyle {\Sigma^*}}^2/M^2},
\label{Sigma2}
\end{eqnarray}
for ${\Xi^*}$:
\begin{eqnarray}
& &
{1\over 5}\; E_2\,L^{4/27}\;M^6
- {5\over 72}\,b\; E_0\,L^{4/27}\;M^2
+ {4\over 9}\,f_s(f_s+2)\,\kappa_v a^2\;L^{28/27}
- {7\over 27}\,f_s(f_s+2)\,m^2_0 a^2\; L^{14/27}{1\over M^2}
\nonumber \\ & &
+{2\over 3}\,m_s\,(f_s+2)\, a\; E_0\,L^{4/27}\;M^2
-{1\over 9}\,m_s\,(2f_s+7)\,m^2_0 a\;L^{-10/27}
=\tilde{\lambda}_{\scriptscriptstyle {\Xi^*}}^2\;
e^{-M_{\scriptscriptstyle {\Xi^*}}^2/M^2},
\label{Xi2}
\end{eqnarray}
and for $\Omega$:
\begin{eqnarray}
& &
{1\over 5}\; E_2\,L^{4/27}\;M^6
- {5\over 72}\,b\; E_0\,L^{4/27}\;M^2
+ {4\over 3}\,f^2_s\,\kappa_v a^2\;L^{28/27}
- {7\over 9}\,f^2_s\,m^2_0 a^2\; L^{14/27}{1\over M^2}
\nonumber \\ & &
+3\,m_s\, f_s\,a\; E_0\,L^{4/27}\;M^2
-{3\over 2}\,m_s\,f_s\,m^2_0 a\;L^{-10/27}
=\tilde{\lambda}_{\scriptscriptstyle \Omega}^2\;
e^{-M_{\scriptscriptstyle \Omega}^2/M^2}.
\label{Omega2}
\end{eqnarray}
In the above equations, $a=-(2\pi)^2\,\langle\bar{u}u\rangle$,
$b=\langle g^2_c\, G^2\rangle$, $\langle\bar{u}g_c\sigma\cdot G
u\rangle=-m_0^2\,\langle\bar{u}u\rangle$,
$\tilde{\lambda}_B=(2\pi)^2\lambda_B$. 
The ratio $f_s=\langle\bar{s}s\rangle/\langle\bar{u}u\rangle
=\langle\bar{s}g_c\sigma\cdot G s\rangle
/\langle\bar{u}g_c\sigma\cdot G u\rangle$ accounts for 
the flavor symmetry breaking of the strange quark.
The four-quark condensate is not well-known and we use the 
factorization approximation 
$\langle\bar{u}u\bar{u}u\rangle=\kappa_v\langle\bar{u}u\rangle^2$ and 
investigate its possible violation via the parameter $\kappa_v$.
The anomalous dimension corrections of the various operators 
are taken into account via the factor $L=\left[{\alpha_s(\mu^2)/
\alpha_s(M^2)}\right] =\left[{\ln(M^2/\Lambda_{QCD}^2)/
\ln(\mu^2/\Lambda_{QCD}^2)}\right]$, where $\mu=500$ MeV is the
renormalization scale and $\Lambda_{QCD}$ is the QCD scale parameter.  
As usual, the excited state contributions are modeled using terms 
on the OPE side surviving $M^2\rightarrow \infty$ under the assumption
of duality, and are represented by the factors 
$E_n(x)=1-e^{-x}\sum_n{x^n/n!}$ with $x=w_B^2/M_B^2$
and $w_B$ an effective continuum threshold. Note that $w_B$
is in principle different for different member of the decuplet and we will 
treat it as a free parameter in the analysis.

These sum rules constitute a complete, up-to-date set of QCD sum rules 
for the decuplet baryon two-point functions under standard 
implementation of the approach.
The reader can find differences in a number of Wilson coefficients 
and in the anomalous dimension corrections,
when comparing with the existing calculations.

\section{Analysis}
\label{ana}

The basic steps of the Monte-Carlo based 
analysis are as follows~\cite{Derek96}.  
Given the uncertainties in the QCD input 
parameters, randomly-selected, Gaussianly-distributed sets are generated,
from which an uncertainty distribution in the OPE can
be constructed.  Then a $\chi^2$ minimization is applied to the sum
rule by adjusting the phenomenological fit parameters.  This is done
for each QCD parameter set, resulting in distributions for
phenomenological fit parameters, from which errors are derived.
Usually, 100 such configurations are sufficient for getting stable
results. We generally select 1000 sets which help resolve more subtle
correlations among the QCD parameters and the phenomenological fit parameters.

The Borel window over which the two sides
of a sum rule are matched is determined by the following two criteria:
a) {\em OPE convergence} --- the highest-dimension-operators
contribute no more than 10\% to the QCD side;
b) {\em ground-state dominance} --- excited
state contributions should not exceed more than 50\% of the 
phenomenological side.
The former effectively establishes a lower limit, the latter an upper limit.
Those sum rules which do not have a valid Borel window under these 
criteria are considered unreliable and therefore discarded.  
The emphasis here is on exploring
the QCD parameter space via Monte Carlo.  The 10\%-50\% criteria are a
reasonable choice that provide a basis for quantitative analysis.
Reasonable alternatives to the criteria are automatically
explored in the Monte-Carlo analysis, as the condensate values and the
continuum threshold would change in each sample.

The QCD input parameters and their uncertainty assignments are 
given as follows.
The quark condensate in standard notation 
is taken as $a=0.52\pm0.05$ GeV$^3$, corresponding to a central value of 
$\langle\bar{u}u\rangle=-(236)^3$ MeV$^3$.
For the gluon condensate,
early estimates from charmonium~\cite{SVZ79} place it at 
$b=0.47\pm0.2$ GeV$^4$, a value commonly used in QCD sum rule 
analysis. But more recent investigations support 
much larger values~\cite{Marrow87,Bert88,Narison95,Ji95}. Here we adopt 
$b=1.2\pm0.6$ GeV$^4$ with 50\% uncertainty.
The mixed condensate parameter is placed at $m^2_0=0.72\pm0.08$ GeV$^2$.
For the four-quark condensate, there are claims of significant 
violation of the factorization hypothesis~\cite{Marrow87,Bert88,Narison95}.
Here we use $\kappa_v=2\pm 1$ and $1\leq \kappa_v \leq 4$.
The QCD scale parameter is restricted to $\Lambda_{QCD}$=0.15$\pm$0.04 GeV.
We find variations of $\Lambda_{QCD}$ have little effects on the results. 
The strange quark mass is taken as $m_s=0.15\pm 0.02$ GeV.
The value of $f_s$ has been determined in~\cite{RRY85,Bely83} and is
given by $f_s=0.83\pm0.05$ after converting to our notation by 
$\gamma=f_s-1$.
These uncertainties are assigned conservatively and in accord with the
state-of-the-art in the literature. 
While some may argue that some
values are better known, others may find that the errors are
underestimated.  In any event, one will learn how the uncertainties in
the QCD parameters are mapped into uncertainties in the
phenomenological fit parameters.

\begin{figure}[p]
\centerline{\psfig{file=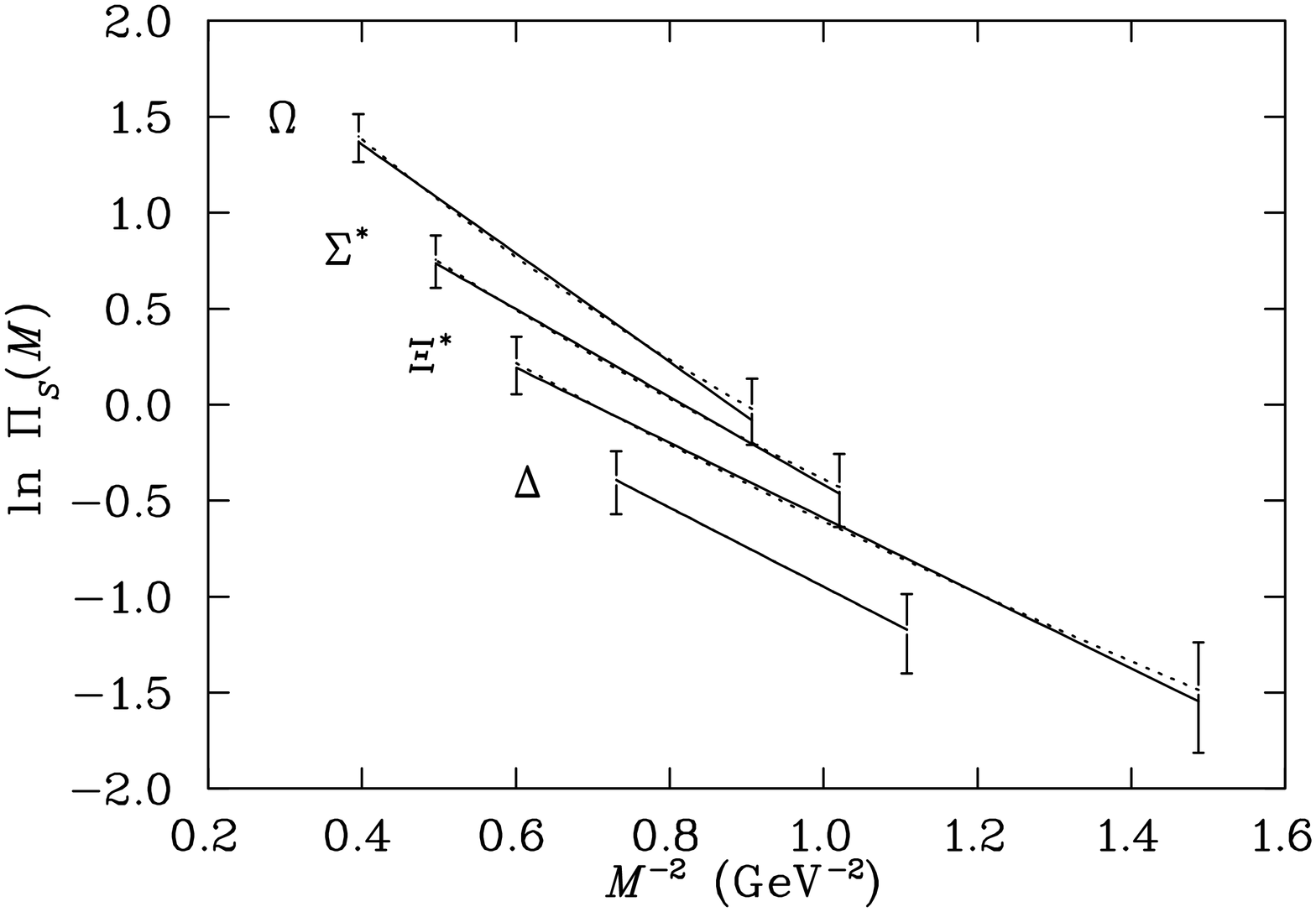,height=8cm,width=12cm,angle=90}}
\vspace{1cm}
\caption{Monte-Carlo fits of the chiral-odd sum rules at the 
structure $g_{\scriptscriptstyle \mu\nu}$.
Each sum rule is searched independently.
The solid line corresponds  to the ground state contribution, 
the dotted line the rest of the contributions (OPE minus continuum).
The error bars are only shown at the two ends for clarity.}
\label{fit1}
\end{figure}

\begin{figure}[p]
\centerline{\psfig{file=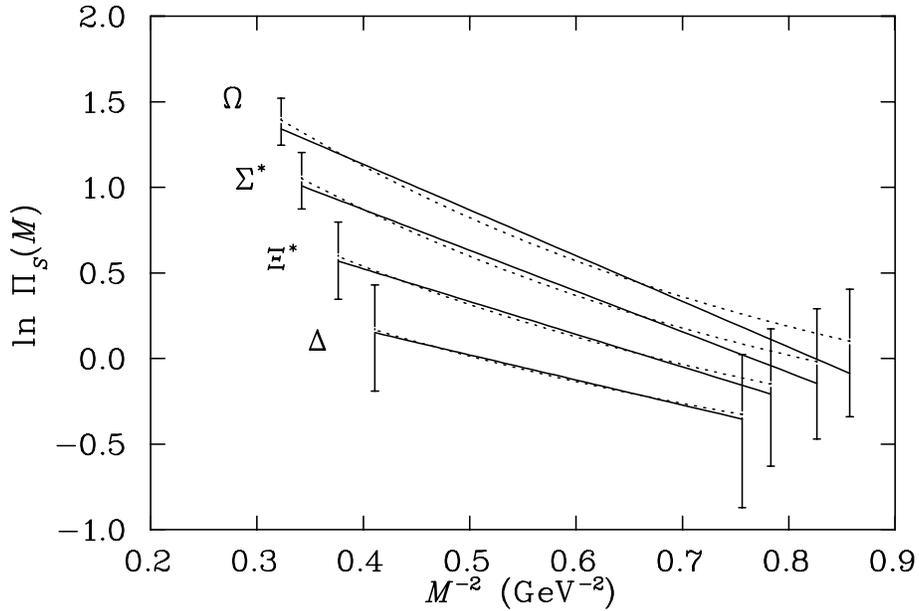,height=8cm,width=12cm,angle=90}}
\vspace{1cm}
\caption{Similar to Fig.~\protect{\ref{fit1}},
but for the chiral-even sum rules (without $\alpha_s$ corrections)
at the structure $g_{\scriptscriptstyle \mu\nu}\hat{p}$.
For better viewing, the curves for each baryon have been shifted 
downward by 0.2 units relative to the previous one.}
\label{fit2}
\end{figure}

In illustrating how well a sum rule works, we choose to plot
the logarithm of the two sides of the sum rule
against the inverse of $M^2$.
In this way, the right-hand side will appear as a straight line
whose slope is $-M_B^2$ and whose intercept with the y-axis gives
a measure of the coupling strength.
The linearity (or deviation from it) of the left-hand side gives 
an indication of OPE convergence and the quality of the continuum model.
The two curves should match in the defined Borel region for a good sum rule.
Fig.~\ref{fit1} shows such a plot for the
chiral-odd sum rules at the structure $g_{\scriptscriptstyle \mu\nu}$.
The resulting fit parameters are given in Table~\ref{tabm1}.
The experimental masses are taken from Particle Data Group~\cite{PDG97}.
Fig.~\ref{fit2} shows a similar plot for the chiral-even sum rules.
The resulting fit parameters are given in Table~\ref{tabm2}.

%
%
%
%
%
\begin{table}[tb]
\caption{Monte-Carlo analysis of the chiral-odd sum rules for the
decuplet. The third column is the percentage contribution of the excited
states to the phenomenological side at the lower end of the Borel region 
(it increases to 50\% at the upper end).
The results are obtained from consideration of 1000 QCD parameter sets.}
\label{tabm1}
\begin{tabular}{ccccccc}
Sum Rule & Region & Cont & $w$ & $\tilde{\lambda}^2$ & Mass & Exp.\\
& (GeV) & (\%) & (GeV) & (GeV$^6$) & (GeV) & (GeV)\\ \hline
$\Delta$
& 0.95 to 1.17  & 31 & 1.65$\pm$ 0.22 & 2.26$\pm$ 0.89 & 1.43$\pm$ 0.12 & 1.232 \\
${\Sigma^*}$
& 0.82 to 1.29  & 8 & 1.80      & 2.83$\pm$ 0.32 & 1.394$\pm$ 0.052& 1.384 \\
${\Xi^*}$
& 0.99 to 1.42  & 13 & 2.00     & 4.32$\pm$ 0.47 & 1.505 $\pm$ 0.037  & 1.533 \\
$\Omega$
& 1.05 to 1.59  & 8 & 2.30      & 7.19$\pm$ 0.75 & 1.676 $\pm$ 0.031& 1.672 \\
\end{tabular}
\end{table}
%
%
\begin{table}[tb]
\caption{Similar to Table~\protect\ref{tabm1}, but for the 
chiral-even sum rules.
The second row in each sum rule shows the results with the leading-order 
$\alpha_s$ corrections included.}
\label{tabm2}
\begin{tabular}{ccccccc}
Sum Rule & Region & Cont & $w$ & $\tilde{\lambda}^2$ & Mass & Exp.\\
& (GeV) & (\%) & (GeV) & (GeV$^6$) & (GeV) & (GeV)\\ \hline
$\Delta$
& 1.15 to 1.56 & 10 & 2.20      & 4.13$\pm$ 0.65 & 1.19$\pm$ 0.13 & 1.232 \\
& 1.15 to 1.39 & 23 & 2.00      & 3.79$\pm$ 0.51 & 1.23 $\pm$ 0.11 &  \\
${\Sigma^*}$
& 1.13 to 1.63  & 6 & 2.40      & 5.63$\pm$ 0.48 & 1.36 $\pm$ 0.13 & 1.384 \\
& 1.13 to 1.47  & 15& 2.20     & 5.49$\pm$ 0.41 & 1.38 $\pm$ 0.11 &  \\
${\Xi^*}$
& 1.10 to 1.71  & 3 & 2.60      & 7.71$\pm$ 0.37 & 1.52 $\pm$ 0.13  & 1.533 \\
& 1.10 to 1.56  & 8 & 2.40      & 7.92$\pm$ 0.43 & 1.53 $\pm$ 0.10 &  \\
$\Omega$
& 1.08 to 1.76  & 2 & 2.70      & 9.16$\pm$ 0.43 & 1.61 $\pm$ 0.12 & 1.672 \\
& 1.08 to 1.61  & 5 & 2.50      & 9.54 $\pm$ 0.53 & 1.61 $\pm$ 0.10 &  \\
\end{tabular}
\end{table}
%

First, let us note that 
the chiral-odd $\Delta$ sum rule~(\ref{Delta1}) is the only one that allows 
a three-parameter search. 
In the other sum rules, there is not enough information 
in the OPE for a three-parameter search.
What happens numerically is that 
the search algorithm return a continuum threshold that is either 
zero or smaller than the mass, which is clearly unphysical.
In order to proceed, we decide to fix the continuum thresholds 
at certain values and perform a two-parameter search on the masses 
and couplings.
By adjusting the continuum thresholds to values 
that reproduce the experimentally known masses,
the current couplings $\tilde{\lambda}_B^2$ 
are left as predictions from the 
self-consistency requirement of the corresponding sum rules.
We have tried a different approach: fixing the masses at known values
and search for the continuum thresholds and couplings. 
This was not successful due to the same reasons as above.
It is satisfying to observe that the obtained continuum thresholds 
in Table~\ref{tabm1} roughly coincide with the first excited state 
in each channel from Particle Data Group~\cite{PDG97}:
$\Delta(1600)$, $\Sigma^*(1840)$, $\Xi^*(1950)$, $\Omega(2250)$.

The predicted $\Delta$ mass lies above the experimental value 
by less than two standard deviations. This small overestimation 
of $M_\Delta$ is a typical result in the QCD sum rule approach. 
For example, in Ref.~\cite{Bely82} it was found 
$M_\Delta\simeq 1.37$ GeV, $\tilde{\lambda}_B^2\simeq 2.3$ GeV$^6$,
and $w$=2.2 GeV; 
and in Ref.~\cite{Derek90} it was found 
$M_\Delta\simeq 1.36$ GeV, $\tilde{\lambda}_B^2\simeq 1.53$ GeV$^6$,
and $w$=1.58 GeV. 
In these works, the traditional value of $b=0.47$ GeV$^4$ was used.
Using this value in the present study yields
$M_\Delta\simeq 1.385$ GeV, $\tilde{\lambda}_B^2\simeq 2.00$ GeV$^6$,
and $w$=1.61 GeV.
The couplings from these studies are consistent with each other.
It is worth mentioning that both $M_\Delta$ and 
$\tilde{\lambda}_B^2$ as determined from the QCD sum rule method 
agree with those from other calculations.
For example, a lattice QCD calculation~\cite{Chu93} gives 
$M_\Delta\simeq 1.43$ GeV, $\tilde{\lambda}_B^2\simeq 2.13$ GeV$^6$;
and an instanton liquid model calculation~\cite{Schafer94} gives
$M_\Delta\simeq 1.43$ GeV, $\tilde{\lambda}_B^2\simeq 1.70$ GeV$^6$.
The reason for the overestimation of the $\Delta$ mass in QCD sum rule 
method remains an open question.
It is likely that the power corrections are large in this sum rule and 
the OPE is not sufficiently convergent.
On the phenomenological side, one notes that the continuum contribution 
in this sum rule is the largest among the decuplet family, 
greater than 30\% in the entire Borel region used.
This also indicates that the continuum model may not be adequate.

Further examination of Fig.~\ref{fit1} and Fig.~\ref{fit2}
reveals that the errors in the chiral-odd sum rules are 
generally smaller than those in the chiral-even sum rules.
More importantly, the convergence of the OPE in the chiral-odd sum rules
are better than that in the chiral-even sum rules 
(as evidenced by the deviation of the dotted line from linearity).
Consequently, the results from the chiral-odd sum rules are 
more reliable than those from the chiral-even sum rules.
This fact is a general feature of baryon two-point functions 
as recently emphasized in Ref.~\cite{Jin97}.
It can be traced to the fact that even and odd parity excited 
states contribute with different signs.
The contributions of positive- and negative-parity excited 
states partially cancel each other in the chiral-odd sum rules,
whereas they add up in the chiral-even sum rules.
Further evidence for the unreliability of chiral-even sum rules
will be discussed below.

The effects of perturbative corrections to first order in $\alpha_s$
were also studied in this work.
These corrections have been calculated in~\cite{Chung84} 
and are given by
\begin{equation}
\left({11\over 9}-{4\over 3}\gamma_E\right){\alpha_s\over \pi}, 
\;\;\;\mbox{and}\;\;\;
\left({17\over 27}-{5\over 6}\gamma_E\right){\alpha_s\over \pi},
\end{equation}
to the dimension-three ($a$) and dimension-five ($m^2_0a$) terms, 
respectively, in the chiral-odd sum rules~(\ref{Delta1}) to~(\ref{Omega1}); 
and 
\begin{equation}
\left({539\over 90}-{1\over 3}\gamma_E\right){\alpha_s\over \pi},
\;\;\;\mbox{and}\;\;\;
-\left({113\over 108}+{22\over 3}\gamma_E\right){\alpha_s\over \pi},
\end{equation}
to the dimension-zero (the identity operator) and dimension-six ($a^2$) terms, 
respectively, in the chiral-even sum rules~(\ref{Delta2}) to~(\ref{Omega2}). 
Here $\gamma_E\simeq 0.58$ is the Euler constant.
At the scale of about 1 GeV$^2$, $\alpha_s/ \pi\simeq 0.12$.
So the leading-order $\alpha_s$ corrections amount to about 
5\% in the chiral-odd sum rules and about 70\% in the chiral-even sum rules. 
Their effects on the spectral parameters are given in Table~\ref{tabm2} 
as a second entry for each of the  chiral-even sum rules.
We find the effects of $\alpha_s$ corrections in the chiral-odd sum rules 
are small and can be safely neglected. 
They were not listed in Table~\ref{tabm1}.
On the other hand, inclusion of $\alpha_s$ corrections 
in the chiral-even sum rules leads to a 
shrinkage of the valid Borel regions and an increase of the continuum
contributions, signs of deterioration of the sum rules. 
Since the identity operator term is closely tied to the 
continuum model, it turns out that the effects of the $\alpha_s$ corrections 
can be compensated by simply shifting down the continuum threshold by 
about 200 MeV.

When considering the uncomfortably large $\alpha_s$ corrections 
in the chiral-even sum rules, the possibility of a significant dimension-two 
power correction from a summation of the perturbative 
series~\cite{Zakharov92,Brown92}, coupled with 
the large uncertainties associated with the four-quark condensate, 
one has to conclude that the QCD side of these sum rules are really 
poorly known. 
The spectral properties extracted from them are most likely unreliable.
Therefore, we caution against the use of these chiral-even sum rules
in favor of the chiral-odd sum rules.

\begin{figure}[th]
\centerline{\psfig{file=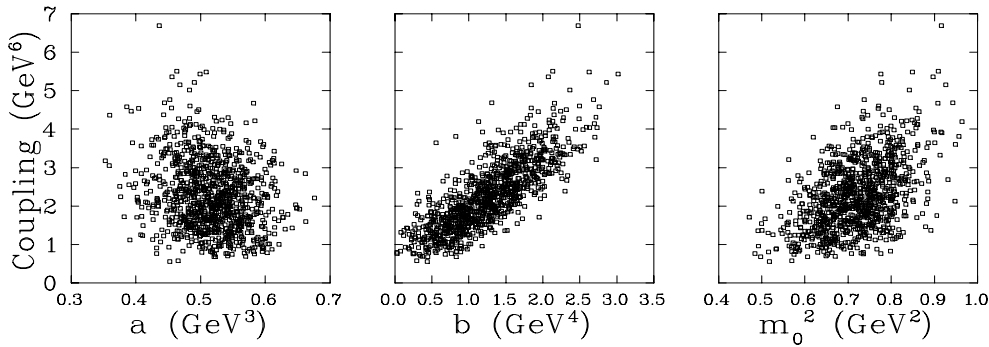,height=4.5cm,width=13cm,angle=90}}
\vspace{1cm}
\caption{Scatter plots showing correlations between $\tilde{\lambda}_\Delta^2$
and the QCD input parameters for the chiral-odd sum rule (\protect\ref{Delta1}).
The results are drawn from 1000 QCD parameter sets. 
The distributions for $M_\Delta$ and $w_\Delta$ have 
qualitatively the same shapes as for $\tilde{\lambda}_\Delta^2$, 
and are not shown here.}
\label{corrqcd_del1}
\end{figure}

Since all the parameters in the Monte-Carlo analysis are
correlated, one can study the correlations between any two parameters
by looking at their scatter plots.  Such plots are useful in 
revealing how a particular sum rule resolves the spectral 
properties. Fig.~\ref{corrqcd_del1} shows the
correlations of vacuum condensates with the $\Delta$ coupling for 
the chiral-odd sum rule~(\ref{Delta1}).  
Similar plots hold for the mass and the continuum threshold and 
are not shown. The uncertainties in the quark condensate reveal 
anti-correlations with the spectral parameters, while those in 
the gluon condensate and the mixed condensate display positive correlations. 
The fact that all of the condensates have significant correlations 
with the phenomenological parameters suggests that all three terms 
in the OPE side of this sum rule play an important role 
in determining the outcome.
This may be an indication that the OPE in this sum rule is not yet 
sufficiently convergent.
The fact that all three phenomenological parameters are correlated 
with the vacuum condensates in the same manner suggests that attempts 
to fine tune the condensates will increase or decrease 
the phenomenological parameters simultaneously in this sum rule.
%
%

\begin{figure}[hb]
\centerline{\psfig{file=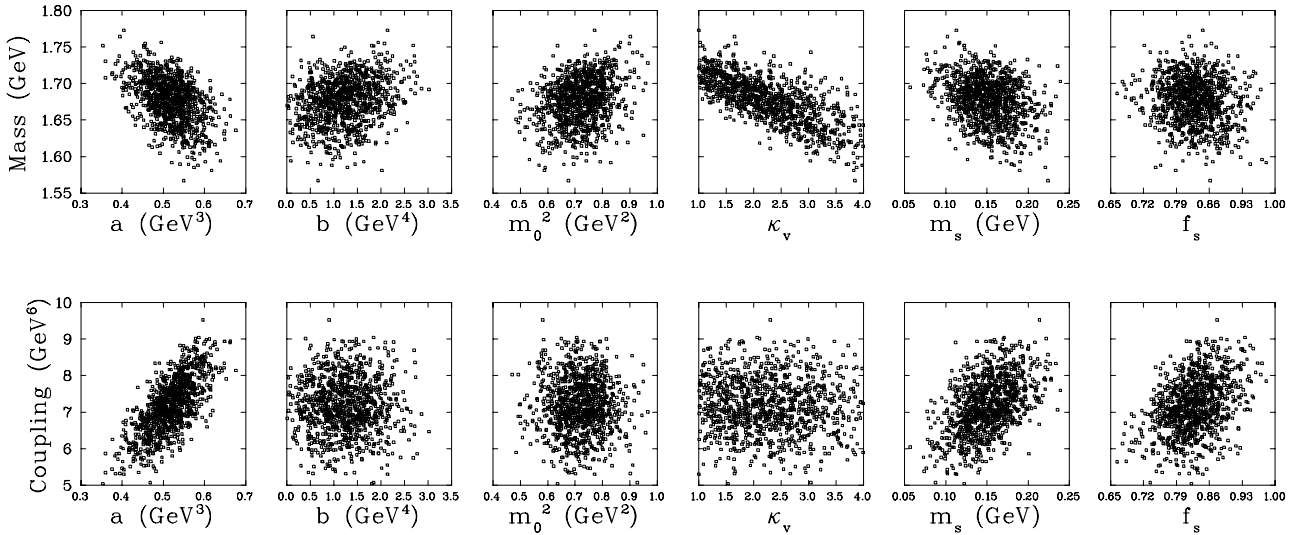,height=7cm,width=17cm,angle=90}}
\vspace{1cm}
\caption{Similar to Fig.~\protect{\ref{corrqcd_del1}}, 
but for the chiral-odd $\Omega$ sum rule (\protect\ref{Omega1}).}
\label{corrqcd_omeg1}
\end{figure}

Fig.~\ref{corrqcd_omeg1} shows similar scatter plots 
for $\Omega$ at the chiral-odd sum rule~(\ref{Omega1}).
Here three more parameters come into play: $\kappa_v$, $m_s$ and $f_s$.
It is interesting to observe that the mass and the coupling have 
different correlation patterns.
They are opposite for the quark condensate.
Similar is true for the strange quark mass, although less pronounced.
The mass is negatively correlated with $\kappa_v$, while the coupling 
has little correlation with it.
The correlations with $b$, $m_0^2$ and $f_s$ appear fairly weak.

We have examined scatter plots for all of the sum rules.
We find that qualitatively, the strange baryon ($\Sigma^*$, $\Xi^*$ and
$\Omega$) sum rules have very similar patterns within the same 
chirality, despite some subtle differences.
The patterns in the chiral-even sum rules are very different from 
those in the chiral-odd sum rules. We will not elaborate further about them.

%
%

In a recent work~\cite{Dey97}, the authors observed that 
baryon current couplings and their masses obey 
a simple cubic scaling law: $\lambda_B \sim M_B^3$, 
across the octet as well as the decuplet.
This result is based on the couplings
extracted from existing QCDSR calculations.
Some theoretical justification was given based on general scaling 
arguments for QCD  and a simple light-cone constituent quark model.
Using our results for the decuplet, we are able to check more carefully 
this claim.

\begin{figure}[tb]
\centerline{\psfig{file=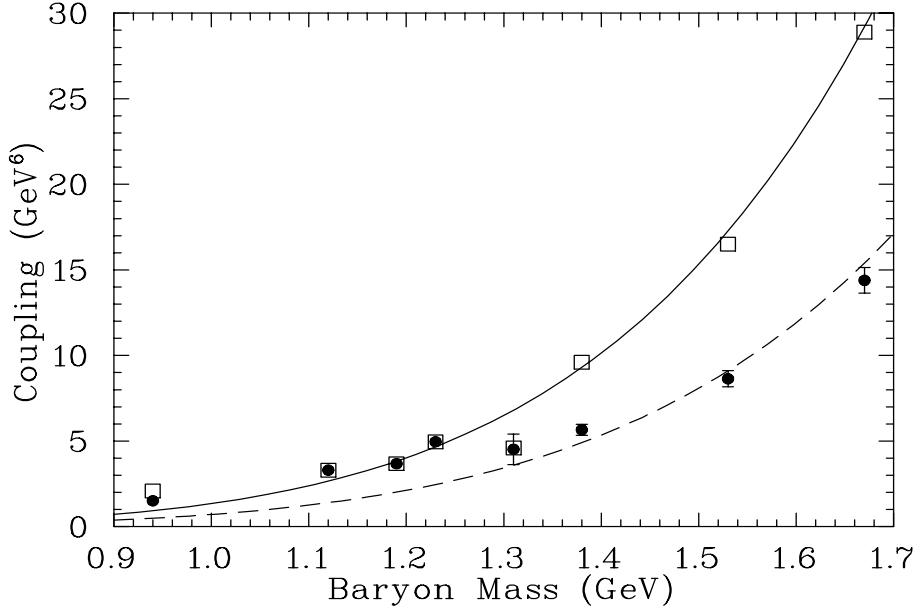,height=8cm,width=12cm,angle=90}}
\vspace{1cm}
\caption{Baryon current coupling ($\tilde{\lambda}_B^2$) versus
the baryon mass($M_B$).
The empty squares are the points extracted by
Ref.~\protect\cite{Dey97}, and the filled circles are from this work.
The solid line represents the fit to the square points:
$\tilde{\lambda}_B^2=(1.16)^2 M_B^6$, as obtained in Ref.~\protect\cite{Dey97}.
The dashed line is the fit to our points: 
$\tilde{\lambda}_B^2=(0.84)^2 M_B^6$.}
\label{scaling}
\end{figure}

In Fig.~\ref{scaling}, following Ref.~\cite{Dey97}, we plot the couplings 
$\tilde{\lambda}_B^2\equiv 2(2\pi)^4 \lambda_B^2$ 
as a function of the masses, using points extracted by 
Ref.~\cite{Dey97} and our points.
The nucleon point is taken from Ref.~\cite{Derek96} which is the 
most recent analysis of the nucleon mass sum rule.
It lies slightly below the point used in Ref.~\cite{Dey97}.
For the octet $\Lambda$, $\Sigma$ and $\Xi$, there exist no new 
analyses to our knowledge, so the same points are used.
The important difference is that our values from Table~\ref{tabm1} 
for the decuplet $\Sigma^*$, $\Xi^*$ and $\Omega$ are
significantly smaller than the ones extracted by Ref.~\cite{Dey97}. 
The $\Delta$ points roughly agree.
This is true even for the values extracted from the 
less reliable chiral-even sum rules in 
Table~\ref{tabm2}, which are slightly larger than those from 
the chiral-odd sum rules. 
Note that we have taken into account a factor of 2 difference in the 
definition of $\tilde{\lambda}_B^2$ in the comparisons.
We attempted to fit the new points with a cubic law and obtained 
$\tilde{\lambda}_B^2=0.71 M_B^6$ compared to their result of
$\tilde{\lambda}_B^2=1.35 M_B^6$.
Notice, however, that there are considerable deviations from the points 
in the new fit, especially the octet strange baryon points.
This calls for a re-examination of the QCD sum rules for the octet 
$\Lambda$, $\Sigma$ and $\Xi$. 
Given the great importance of a scaling law between baryon current 
couplings and their masses and its phenomenological consequences,  
more investigations are clearly needed to resolve the 
deviations, perhaps coupled with other methods such as lattice QCD.

%
%

\section{Conclusion}
\label{con}

We have re-derived and re-analyzed in detail the QCD sum rules for 
decuplet baryon two-point functions using a comprehensive 
Monte-Carlo based procedure.
Predictions were obtained for the current couplings with 
an accuracy on the order of 10\% for $\Sigma^*$, $\Xi^*$ and $\Omega$,
and 40\% for $\Delta$ (see Table~\ref{tabm1}), using realistic 
error estimates for the QCD input parameters.
The results are useful in evaluating matrix elements of 
decuplet baryons in the QCD sum rule approach.
Our calculations confirmed the general feature that chiral-odd sum rules 
are more reliable than chiral-even sum rules as far as 
baryon two-point functions are concerned.
Correlations between the QCD input parameters and the phenomenological
parameters  are studied by way of scatter plots. Some insights 
are gained on how a particular sum rule resolves the spectral properties.
Finally, the couplings obtained in the present study
are used to check a possible cubic scaling law between 
baryon couplings and their masses, as recently claimed in 
Ref.~\cite{Dey97}. We find significant reduction in the 
scaling constant and possible deviations from the cubic law.
More studies are needed to clarify this important issue.

\acknowledgements
It is a pleasure to thank D.B. Leinweber for sharing his Monte-Carlo
analysis program and for helpful discussions.
This work was supported in part by 
U.S. DOE under Grant DE-FG03-93DR-40774.

\end{document}